\documentclass[reprint, aps, prb, amsmath, amssymb, superscriptaddress, longbibliography]{revtex4-2}
\usepackage{bm}
\usepackage{graphicx}
\usepackage{comment}
\usepackage{float}
\usepackage{multirow}
\usepackage{braket}
\usepackage[colorlinks, linkcolor= blue, citecolor = blue, urlcolor=blue]{hyperref}

\def\be{\begin{equation}}
\def\ee{\end{equation}}
\def \bea{\begin{eqnarray}}
\def \eea{\end{eqnarray}}
\def \nn{\nonumber}

\begin{document}
\title{Band Geometry Induced Third-Harmonic Generation}
\author{Sanjay Sarkar}
\thanks{S.S. and D.M. contributed equally.}
\affiliation{Department of Physics, Indian Institute of Technology Kanpur, Kanpur 208016, India}
\author{Debottam Mandal}
\thanks{S.S. and D.M. contributed equally.}
\affiliation{Department of Physics, Indian Institute of Technology Kanpur, Kanpur 208016, India}
\affiliation{Dipartimento di Fisica "E. R. Caianiello", Universita‘ di Salerno, IT-84084 Fisciano (SA), Italy}
\author{Amit Agarwal}
\email{amitag@iitk.ac.in}
\affiliation{Department of Physics, Indian Institute of Technology Kanpur, Kanpur 208016, India}

\begin{abstract}
Third-harmonic generation (THG) is a key nonlinear optical process for ultrafast imaging, terahertz (THz) signal generation, and symmetry-sensitive probes, often dominating in centrosymmetric materials where lower-order responses vanish. Yet, the role of band geometry, Fermi surface effects, and disorder in enabling large and tunable THG remains poorly understood. Here, we develop a finite-frequency quantum kinetic theory of THG based on the density matrix formalism, deriving the third-harmonic conductivity tensor. Our framework isolates five distinct band-geometric contributions to interband and intraband processes, separates Fermi sea from Fermi surface terms, and incorporates disorder effects phenomenologically. We further provide a complete symmetry classification of THG for all 122 magnetic point groups. Applying the theory to the spin-split altermagnet RuO$_2$, we trace its THG response to specific geometric terms. These results establish a predictive foundation for designing materials with enhanced and tunable THG in the finite-frequency regime.
\end{abstract}

\maketitle

\section{Introduction}
Since the advent of the laser, nonlinear optics~\cite{Boyd2008_optics_book} has remained central to photonics research, driving advances in photovoltaics~\cite{Orenstein2021_annual,QMa2023_photocurr}, magnetic memory writing, and telecommunication technologies~\cite{Zhang2021_THzdetect,Jia2023_telecom}. Specifically, third-harmonic generation (THG) has emerged as a powerful nonlinear optical process, enabling ultrafast imaging, THz signal generation, and symmetry-sensitive probes in quantum materials~\cite{Rogalski2019_THzoutlook,Jia2023_telecom,Priessnitz2024_pdcoo2,Reinhoffer2024_strongTHzthg,Maleki2025_strategyTHz,Karvonen2017_grain}. THG occurs when incident light induces a polarization oscillating at three times the driving frequency ($\omega$), producing photons or responses at $3\omega$. While higher harmonics often coexist, THG dominates in systems where symmetry forbids lower-order responses, such as centrosymmetric materials where second-harmonic susceptibility vanishes~\cite{Lai2021_bcp_third}.

Graphene has emerged as a prominent THG platform~\cite{Liu2018_book_grphn,Shang2021_grphn_rev,Vermeulen2022_perspect} due to its linear dispersion, broad bandwidth, and high carrier mobility. 
Experiments report strong THG in monolayer graphene, with the magnitudes of the nonlinear susceptibility varying by the order of $10^5$ to $10^6$~\cite{Nardeep2013_third,Soavi2018_broadbandthg,Jiang2018_gate,Inukai2023_cmplxthg,DiGaspare2023_thzslg} in several reports. This large variation arises from factors such as doping, photon energy, and laser power~\cite{Jiang2018_gate,Soavi2018_broadbandthg,Soavi2019_hot}. Large THG signals often arise from multiphoton resonances involving interband and intraband transitions~\cite{Cheng2014_graphene_optical,Mikhailov2014_graphene,Glazov2014_graphene_high,Cheng2015_graphene_pheno,Rostami2016_diagram,Cheng2019_intra,Maleki2025_strategyTHz}, and can be further amplified by carrier relaxation~\cite{Cheng2015_graphene_pheno}. Beyond graphene, THG has been observed in a range of materials, including 1D~\cite{Zheng2022_1ddf} and 3D Dirac fermions~\cite{Cheng2020_acs_3d}, bilayer graphene~\cite{McGouran2016_THzblg}, twisted bilayers~\cite{Ha2021_thgtblg}, heterostructures~\cite{Calafell2021_thgrphn}, Dirac semimetals~\cite{Ullah2020_thgdsm}, topological insulators~\cite{Xiang2021_thgti}, van der Waals layered materials~\cite{Youngblood2017_layerthgacs,Zhu2025_bp_mid_infra}, transition metal oxides~\cite{Priessnitz2024_pdcoo2,Reinhoffer2024_strongTHzthg}, and dichalcogenides~\cite{Woodward2017_microscopy,Saynatjoki2017_thgmos2}.

In parallel, recent work has revealed deep links between nonlinear optics and the geometric structure of Bloch wavefunctions~\cite{Morimoto2016a,Nagaosa2017,Ahn2021_Riemannian,Sinha2022,Torma2023_essay,Liu2024_geo_rev,Adak2024,Jiang2025_reveal_geo,dutta2025nonlinear}. The Berry connection polarizability (BCP) tensor, for instance, can drive third-harmonic Hall responses in bulk T$_d$-MoTe$_2$~\cite{Lai2021_bcp_third}, generating Hall currents in nonmagnetic centrosymmetric systems through electric-field-induced Berry curvature corrections. Similar BCP-driven effects have been observed in Dirac semimetal Cd$_3$As$_2$~\cite{Zhao2023_gate_BCDP}. The dissipative component of the BCP, tied to the quantum metric quadrupole, has been detected in MnBi$_2$Te$_4$~\cite{Li2024_geoquad} and WTe$_2$~\cite{Liu2025_qmquad} at room temperature. Berry curvature quadrupoles, related to the imaginary quantum geometric tensor, yield extrinsic Hall currents in MnBi$_2$Te$_4$~\cite{Li2024_geoquad} and the kagome antiferromagnet FeSn~\cite{Sankar2024_expbcq}.

Despite these advances, most theoretical studies of geometry-driven third-order responses~\cite{HLiu2022_BCP_prb,Ye2022_orbital,Xiang2023_prb_intrinsic_third,Nag2023_third_ti,Zhu2023_third_mtsm,YGao2023,Zhang2023_multipole,Sorn2024_octupolar,fang_PRL2024_quan,Mandal2024_quangeo} focus on the dc limit ($\omega \to 0$). 
The finite-frequency regime, crucial for THz and GHz technologies, remains largely unexplored. Specifically, there is a lack of a predictive framework that captures the interplay of band geometry, disorder, and Fermi surface effects in designing materials with large, tunable THG.

Here, we present a systematic finite-frequency quantum kinetic theory of THG based on the density matrix formalism. Our framework derives the third-harmonic conductivity tensor from the optically induced polarization, isolates five key geometric contributions to intraband and interband processes, and cleanly separates the Fermi sea from the Fermi surface terms. Disorder is incorporated explicitly through relaxation-time effects. In addition, we provide a full symmetry classification of THG across all 122 magnetic point groups. As a case study, we compute the THG response in the spin-split altermagnet RuO$_2$ and trace its origin to specific band-geometric terms. 

The paper is organized as follows. In Section~\ref {theo_sec}, we develop the general theory of THG, while Section~\ref {polar} analyzes its dependence on linearly polarized light. Section~\ref{THG_graphene} benchmarks the developed framework for graphene. Section~\ref{model_sec} applies it to the spin-split altermagnet RuO$_2$. Section~\ref{sym_sec} presents the crystalline symmetry classification. Section~\ref{discussion} provides a broader discussion, and we conclude in Section~\ref{disc_sec} with a summary and outlook.


\section{Theory of third-harmonic Generation\label{theo_sec}}
In this section, we present a versatile quantum kinetic framework based on the density matrix formalism to calculate third-order nonlinear optical conductivities, which give rise to THG. Our focus is on the response of a quantum system driven by a monochromatic electric field of frequency $\omega$, and we derive the resulting third-order current that oscillates at frequency $3\omega$. This formalism captures both interband and intraband contributions on equal footing and provides a microscopic basis for interpreting THG in terms of the underlying band geometry of Bloch electrons. 

\subsection{Density matrix framework} 
We consider a crystalline solid subjected to a time-dependent electric field, described by the total Hamiltonian $\hat{\cal H} = \hat{\cal H}_0 + \hat{\cal H}_{\rm E}$.  $\hat{H}_0$ is the unperturbed Bloch Hamiltonian and $\hat{H}_{\rm E}$ represents the coupling to the external field. There are two commonly used formulations for incorporating the electric field: the velocity gauge~\cite{Moss1990, Sipe1993} and the length gauge~\cite{ Aversa1995_prb_length, Sipe2000,virk_PRB2007_semi}. These two formulations are related through a time-dependent unitary transformation and yield the same physical observables when a complete electronic basis is used, to all orders of perturbation theory~\cite{ventura_PRB2017_gauge}. In our paper, we adopt the length gauge. Using the dipole approximation, the interaction Hamiltonian is expressed as 
\be
\hat{\cal H}_{\rm E} = -e{\bm E}(t) \cdot \hat{\bm r} ~.
\ee
Here, $e$ is the electronic charge, $\hat{\bm r}$ is the position operator, and ${\bm E}(t)$ is the time-dependent electric field. We work in the basis of Bloch states \( \ket{n\bm{k}} \) such that \( \hat{\cal H}_{0}\ket{n\bm{k}} = \hbar\omega_{n}\ket{n\bm{k}} \). A key aspect of using the length gauge framework is to decompose the position operator into intraband ($\hat{\bm r}_i$) and interband ($\hat{\bm r}_e$) contributions~\cite{Blount1962_band}, 
\bea \label{position_matrix}
    \bra{n\bm{k}}\hat{\bm r}_i\ket{m\bm{k'}} &=& \delta_{nm}[\delta(\bm{k}-\bm{k'})\bm{\mathcal{R}}_{nn} + i \nabla_{\bm{k}}\delta(\bm{k}-\bm{k'})]~, \nn\\
    \\
    \bra{n\bm{k}}\hat{\bm r}_e\ket{m\bm{k'}} &=& (1 - \delta_{nm})\delta(\bm{k}-\bm{k'})\bm{\mathcal{R}}_{nm}~.
\eea
Here, $ \bm{\mathcal{R}}_{nm} = i \langle u_n | \partial_{\bm{k}} u_m \rangle $ with $n \neq m$ defines the interband Berry connection, and $ \bm{\mathcal{R}}_{nn} = i \langle u_n | \partial_{\bm{k}} u_n \rangle $ is the intraband Berry connection. A known technical challenge in using this approach is the singular behaviour of the intraband position operator. However, this difficulty is avoided when the intraband position operator appears inside a commutator, which remains well defined~\cite{Aversa1995_prb_length}. Following Ref.~\cite{Aversa1995_prb_length}, the commutator of $\hat{\bm r}_i$ with any simple operator $S$ satisfies, 
\begin{equation}\label{commute}
[\hat{\bm r_i}, S]_{nm}=(S_{nm}){;\bm{k}} = \frac{\partial S_{nm}}{\partial \bm{k}} - i S_{nm} (\bm{\mathcal{R}}_{nn} - \bm{\mathcal{R}}_{mm})~. 
\end{equation} 
This covariant derivative identity plays an important role in our calculations and will be used extensively throughout the manuscript.

To calculate the nonlinear optical response, we evaluate the nonequilibrium density matrix using the quantum Liouville equation (QLE)~\cite{culcer_PRB2017_inter}, 
\begin{equation}\label{QLE} 
   \dfrac{d\rho}{dt}+\frac{i}{\hbar}[\mathcal{H},\rho]=0~,
\end{equation}
which governs the time evolution of the full density matrix $\rho$ under the total Hamiltonian $\hat{\cal H} = \hat{\cal H}_0 + \hat{\cal H}_{\mathrm{E}}$. 
 By decomposing the position operator in $\hat{\cal H}_{\mathrm{E}}$ into intraband and interband components, $\hat{\bm r}=\hat{\bm r}_i+\hat{\bm r}_e$, we obtain the equation for the density operator matrix elements, 
\begin{equation}\label{eq:7} 
    \dfrac{d\rho_{nm}}{dt}+\frac{i}{\hbar}[\hat{\cal H}_{0},\rho]_{nm}+\frac{\rho_{nm}}{\tau}=\frac{i\textit{e}}{\hbar}\bm{E}(t)\cdot[\hat{\bm r}_e+\hat{\bm r}_i,\rho]_{nm}~.
\end{equation}
To incorporate disorder, we adopt the standard prescription of adiabatically switching on the electric field, 
 ${\bm E}(t) \to {\bm E}(t) e^{-i(\omega+i\eta) t}$, with $\eta=1/\tau$ and $\tau$ is the phenomenological relaxation time. This introduces scattering by making the density matrix disorder-dependent. The advantage of this approach is that the driving electric field in Eq.~\eqref{eq:7} remains harmonic, with ${\bm E}(t) = {\bm E} e^{-i\omega t}+c.c$. Using the commutator identity from Eq.~\eqref{commute}, we rewrite Eq.~\eqref{eq:7} as 
\begin{eqnarray}
\dfrac{d\rho_{nm}}{dt}&+&\frac{i}{\hbar}[\hat{\cal H}_{0},\rho]_{nm}+\frac{\rho_{nm}}{\tau}= \frac{i\textit{e}}{\hbar}\bm{E}(t)\cdot(\rho_{nm})_{;\bm{k}}\nn\\ 
& + &  \frac{i\textit{e}}{\hbar}\bm{E}(t) \sum_{l}(\bm{\mathcal{R}}_{nl}\rho_{lm}-\rho_{nl}\bm{\mathcal{R}}_{lm})  \label{eq_dm_gen}~.  
\end{eqnarray}
  The first term on the right-hand side arises from intraband processes involving the covariant derivative of the density matrix, while the second term captures interband coherence via $\bm{\mathcal{R}}_{nl}$. To solve Eq.~\eqref{eq_dm_gen}, we adopt a perturbative approach by expanding the density matrix in powers of the electric field, $\rho = \sum_i \rho^i$, with $i=0,1,2,...$ and $\rho^i \propto E^i$. This yields a recursive relation for the $N$th-order term \cite{sarkar_arxv_2025}, 
\be
 \rho^{(N)}_{nm}(t)=\frac{i \textit{e}}{\hbar}\int_{-\infty}^{t} dt^\prime e^{i(\omega_{nm}-i\frac{N}{\tau})t^\prime}\bm{E}(t^\prime)~[\bm{R}_{e}^{(N-1)}+\bm{R}_{i}^{(N-1)}]\label{dm_nth}~.
\ee
Here, $\hbar\omega_{nm}=\hbar(\omega_n - \omega_m)$ is the energy difference between bands $n$ and $m$ at a given ${\bm k}$ point and the interband and intraband polarization matrices at the $({N-1}){th}$ order are defined as
\begin{equation}\label{R_e}
    \bm{R}_{e}^{(N-1)}=\sum_{l}[\bm{\mathcal{R}}_{nl}\rho_{lm}^{(N-1)}-\rho_{nl}^{(N-1)}\bm{\mathcal{R}}_{lm}]~, 
\end{equation}%
and
\begin{equation}\label{R_i}
    \bm{R}_{i}^{(N-1)}=[\bm{r}_i,\rho^{(N-1)}]_{nm}=i[\rho_{nm}^{(N-1)}]_{;\bm{k}}~.
\end{equation}
By evaluating Eq.~\eqref{dm_nth} for $N = 1, 2, 3$, we obtain the first-, second-, and third-order density matrices, respectively. In equilibrium (absence of external fields), we assume the system to be in the ground state of $\hat{\cal H}_0$. Thus, the zeroth-order density matrix is $\rho_{nm}^{(0)} = f_n \delta_{nm}$, where $f_n = \left[ 1 + e^{\beta (\epsilon_n - \mu)} \right]^{-1}$ is the Fermi-Dirac distribution with $\beta = 1/(k_B T)$ and $\mu$ being the chemical potential. A detailed derivation of the nonequilibrium density matrix up to third-order in the electric field is presented in Appendix~\ref{App_A}. Having obtained the density matrix, we now calculate the THG response. 

\begin{table*}[t]
	\centering
 \caption{The Fermi sea contributions to the third-order interband optical conductivity, $\sigma_{abcd}^{(3e);{\rm Sea}}(-3\omega; \omega, \omega,\omega) = e^4/\hbar^3 \sum_{\bm k} \tilde{\sigma}_{abcd}^{(3e);{\rm Sea}}({\bm k})$. For simplicity, we define the complex frequencies $\tilde{\omega}=\omega+i/\tau$, $2\tilde{\omega}=2\omega + 2i/\tau$, $3\tilde{\omega}=3\omega+3i/\tau$, with $\tau$ being the relaxation time. We denote the energy bands using $\epsilon_n ({\bm k}) = \hbar \omega_n$, and the corresponding a-component of the band velocity as $ \hbar \omega_{n;a} = v_n^a = \partial_{a} \epsilon_n ({\bm k})$ with $\partial_a \equiv \frac{\partial}{\partial  k_a}$. We denote the energy difference between the pair of bands involved in optical transition at a given ${\bm k}$ point by $\hbar\omega_{nm}=\epsilon_n ({\bm k})- \epsilon_m({\bm k})$, and the corresponding velocity injection is captured by $\omega_{nm;a} = \partial_{a} \omega_{nm}$. In the table, $f_{nm} = f_n - f_m$ with $f_i$ being the Fermi function for the $i-$th band. $\Omega$ is the Berry curvature, ${\cal G}$ is the quantum metric, $\Gamma$ is the metric connection, $\tilde{\Gamma}$ is the symplectic connection and $D$ is the second-order connection. The last three columns indicate which band geometric contributions are finite in the presence of either inversion ($\mathcal P$), time-reversal ($\mathcal T$), or combined ${\mathcal P}{\mathcal T}$ symmetry.
 }
	{
		\begin{tabular}{c@{\hskip 1cm} c@{\hskip 1cm} c@{\hskip 1cm} c @{\hskip 1cm}c }
			\hline \hline
			\rule{0pt}{3ex}
			  Conductivity & Integrand &  $\mathcal{P}$ &  $\mathcal{T}$ &  ${\mathcal P}{\mathcal T}$ \\ [2ex]
			\hline \hline
			\rule{0pt}{3ex}
			 $\tilde{\sigma}_{abcd,1}^{(3e);{\rm Sea}}(\bm{k})$ & $\frac{3\tilde{\omega}}{\tilde{\omega}}\sum_{nm}\frac{(\mathcal{G}_{mn}^{bc}-i\Omega_{mn}^{bc}/2)\Omega_{nm}^{ad}}{(\omega_{nm}-3\tilde{\omega})(\omega_{nm}-\tilde{\omega})} f_{mn}$ & $\mathcal{G}$, $\Omega$ &$\Omega$ &$0$  \\ [2ex]
			 $\tilde{\sigma}_{abcd,2}^{(3e);{\rm Sea}}(\bm{k})$ & $ -i3\tilde{\omega}\sum_{nm}\frac{({\rm Re}[D_{nm}^{abcd}]+i{\rm Im}[D_{nm}^{abcd}])}{(\omega_{nm}-3\tilde{\omega})(\omega_{nm}-2\tilde{\omega})(\omega_{nm}-\tilde{\omega})} f_{mn}$  &$\operatorname{Re}[D]$, $\operatorname{Im}[D]$ & $\operatorname{Re}[D]$ & $\operatorname{Re}[D]$   \\[2ex]
			 $\tilde{\sigma}_{abcd,3}^{(3e);{\rm Sea}}(\bm{k})$ & 	$ i3\tilde{\omega}\sum_{nm}\frac{\omega_{nm;d}(\Gamma_{nm}^{acb}-i\tilde{\Gamma}_{nm}^{acb})}{(\omega_{nm}-3\tilde{\omega})(\omega_{nm}-2\tilde{\omega})(\omega_{nm}-\tilde{\omega})^2}f_{mn}$  & $\Gamma$, $\tilde{\Gamma}$ & $\Gamma$ & $\Gamma$ \\ [2ex]
			 $\tilde{\sigma}_{abcd,4}^{(3e);{\rm Sea}}(\bm{k})$ &  $ i3\tilde{\omega}\sum_{nm}\frac{\omega_{nm;c}(\Gamma_{nm}^{adb}-i\tilde{\Gamma}_{nm}^{adb})}{(\omega_{nm}-3\tilde{\omega})(\omega_{nm}-2\tilde{\omega})(\omega_{nm}-\tilde{\omega})^2}f_{mn}$   & $\Gamma$, $\tilde{\Gamma}$ & $\Gamma$ & $\Gamma$ \\[2ex]
			 $\tilde{\sigma}_{abcd,5}^{(3e);{\rm Sea}}(\bm{k})$ & $ i3\tilde{\omega}\sum_{nm}\frac{\omega_{nm;cd}(\mathcal{G}_{nm}^{ab}-i\Omega_{nm}^{ab}/2)}{(\omega_{nm}-3\tilde{\omega})(\omega_{nm}-2\tilde{\omega})(\omega_{nm}-\tilde{\omega})^2}f_{mn}$  & $\mathcal{G}$, $\Omega$ & $\mathcal{G}$ & $\mathcal{G}$  \\[2ex]
			 $\tilde{\sigma}_{abcd,6}^{(3e);{\rm Sea}}(\bm{k})$ & $ -i3\tilde{\omega}\sum_{nm}\frac{2\omega_{nm;c}\omega_{nm;d}(\mathcal{G}_{nm}^{ab}-i\Omega_{nm}^{ab}/2)}{(\omega_{nm}-3\tilde{\omega})(\omega_{nm}-2\tilde{\omega})(\omega_{nm}-\tilde{\omega})^3}f_{mn}$   & $\mathcal{G}$, $\Omega$ & $\mathcal{G}$ & $\mathcal{G}$ 
             \\[2ex]	
             $\tilde{\sigma}_{abcd,7}^{(3e);{\rm Sea}}(\bm{k})$ & $ i3\tilde{\omega}\sum_{nm}\frac{\omega_{nm;d}(\Gamma_{nm}^{acb}-i\tilde{\Gamma}_{nm}^{acb})}{(\omega_{nm}-3\tilde{\omega})(\omega_{nm}-2\tilde{\omega})^2(\omega_{nm}-\tilde{\omega})}f_{mn}$   & $\Gamma$, $\tilde{\Gamma}$ & $\Gamma$ & $\Gamma$
             \\[2ex]		
             $\tilde{\sigma}_{abcd,8}^{(3e);{\rm Sea}}(\bm{k})$ & $ -i3\tilde{\omega}\sum_{nm}\frac{\omega_{nm;c}\omega_{nm;d}(\mathcal{G}_{nm}^{ab}-i\Omega_{nm}^{ab}/2)}{(\omega_{nm}-3\tilde{\omega})(\omega_{nm}-2\tilde{\omega})^2(\omega_{nm}-\tilde{\omega})}f_{mn}$   & $\mathcal{G}$, $\Omega$ & $\mathcal{G}$ & $\mathcal{G}$ 
             \\[2ex]
             
						\hline \hline
	\end{tabular}}
	\label{table_1}
\end{table*}
\begin{table*}[t]
	\centering
 \caption{The Fermi sea contribution of the third-order intraband optical conductivity, $\sigma_{abcd}^{(3i);{\rm Sea}}(-3\omega; \omega, \omega,\omega) = e^4/\hbar^3 \sum_{\bm k} \tilde{\sigma}_{abcd}^{(3i);{\rm Sea}}({\bm k})$. The notation used for different quantities is the same as that defined in Table~\ref{table_1}.}
	{
		\begin{tabular}{c@{\hskip 1cm} c@{\hskip 1cm} c@{\hskip 1cm} c@{\hskip 1cm} c }
			\hline \hline
			\rule{0pt}{3ex}
			  Conductivity & Integrand & $\mathcal{P}$ & $\mathcal{T}$ & $\mathcal{PT}$ \\ [2ex]
			\hline \hline
			\rule{0pt}{3ex}
            $\tilde{\sigma}_{abcd,1}^{(3i);{\rm Sea}}(\bm{k})$ & 	$\frac{i}{6\tilde{\omega}^2}\sum_{nm}\omega_{nm;a}\omega_{nm;d}\frac{(\mathcal{G}_{nm}^{cb}-i\Omega_{nm}^{cb}/2)f_{mn}}{(\omega_{nm}-\tilde{\omega})^2}$ & $\mathcal{G}$, $\Omega$ &$\mathcal{G}$ &$\mathcal{G}$
            \\ [2ex]
            $\tilde{\sigma}_{abcd,2}^{(3i);{\rm Sea}}(\bm{k})$ & 	$-\frac{i}{6\tilde{\omega}^2}\sum_{nm}\omega_{nm;a}\frac{(\Gamma_{nm}^{cdb}-i\tilde{\Gamma}_{nm}^{cdb})f_{mn}}{(\omega_{nm}-\tilde{\omega})}$  & $\Gamma$, $\tilde{\Gamma}$ & $\Gamma$ & $\Gamma$
            \\ [2ex]
             $\tilde{\sigma}_{abcd,3}^{(3i);{\rm Sea}}(\bm{k})$ & 	$-\frac{i}{6\tilde{\omega}^2}\sum_{nm}\omega_{nm;a}\frac{(\Gamma_{mn}^{bdc}-i\tilde{\Gamma}_{mn}^{bdc})f_{mn}}{(\omega_{nm}-\tilde{\omega})}$ & $\Gamma$, $\tilde{\Gamma}$ & $\Gamma$ & $\Gamma$
             \\ [2ex] 
			 $\tilde{\sigma}_{abcd,4}^{(3i);{\rm Sea}}(\bm{k})$ & $ -\frac{i}{3\tilde{\omega}} \sum_{nm}\frac{\omega_{nm;a}\omega_{nm;c}}{(\omega_{nm}-2\tilde{\omega})}\frac{(\mathcal{G}_{nm}^{db}-i\Omega_{nm}^{db}/2)f_{mn}}{(\omega_{nm}-\tilde{\omega})^2}$    & $\mathcal{G}$, $\Omega$ & $\mathcal{G}$ & $\mathcal{G}$    \\[2ex]
              $\tilde{\sigma}_{abcd,5}^{(3i);{\rm Sea}}(\bm{k})$ & $ \frac{i}{3\tilde{\omega}} \sum_{nm}\frac{\omega_{nm;a}}{(\omega_{nm}-2\tilde{\omega})}\frac{(\Gamma_{nm}^{dcb}-i\tilde{\Gamma}_{nm}^{dcb})f_{mn}}{(\omega_{nm}-\tilde{\omega})}$    & $\Gamma$, $\tilde{\Gamma}$ & $\Gamma$ & $\Gamma$ 
              \\[2ex]
			 $\tilde{\sigma}_{abcd,6}^{(3i);{\rm Sea}}(\bm{k})$ &  $ - \frac{1}{2\tilde{\omega}}\sum_{nm} \Omega_{nm}^{ad}\frac{(\mathcal{G}_{nm}^{cb}-i\Omega_{nm}^{cb}/2)f_{mn}}{(\omega_{nm}-\tilde{\omega})}$  & $\mathcal{G}$, $\Omega$ & $\Omega$ & 0\\[2ex]
			 $\tilde{\sigma}_{abcd,7}^{(3i);{\rm Sea}}(\bm{k})$ & $i\sum_{nm}\frac{ \omega_{nm;c}}{(\omega_{nm}-2\tilde{\omega})}\frac{(\Gamma_{mn}^{bad}-i\tilde{\Gamma}_{mn}^{bad})f_{mn}}{(\omega_{nm}-\tilde{\omega})^2}$    & $\Gamma$, $\tilde{\Gamma}$ & $\Gamma$ & $\Gamma$ 
             \\[2ex]	
             $\tilde{\sigma}_{abcd,8}^{(3i);{\rm Sea}}(\bm{k})$ & $-i\sum_{nm}\frac{1}{(\omega_{nm}-2\tilde{\omega})}\frac{\partial_a\left(\Gamma_{mn}^{dcb}+i\tilde{\Gamma}_{mn}^{dcb}\right)f_{mn}}{(\omega_{nm}-\tilde{\omega})}$   & $\Gamma$, $\tilde{\Gamma}$ & $\Gamma$ & $\Gamma$ 
             \\[2ex]	
             $\tilde{\sigma}_{abcd,9}^{(3i);{\rm Sea}}(\bm{k})$ & $i\sum_{nm}\frac{1}{(\omega_{nm}-2\tilde{\omega})}\frac{\left(\operatorname{Re}[D_{mn}^{dbac}]-i\operatorname{Im}[D_{mn}^{dbac}]\right)f_{mn}}{(\omega_{nm}-\tilde{\omega})}$   &$\operatorname{Re}[D]$, $\operatorname{Im}[D]$ & $\operatorname{Re}[D]$ & $\operatorname{Re}[D]$   \\[2ex]	
						\hline \hline
	\end{tabular}}
	\label{table_2}
\end{table*}

\begin{table*}[t]
	\centering
 \caption{The Fermi surface contributions to the third-order interband optical conductivity, $\sigma_{abcd}^{(3e);{\rm Surface}}(-3\omega; \omega, \omega,\omega) = e^4/\hbar^3 \sum_{\bm k} \tilde{\sigma}_{abcd}^{(3e);{\rm Surface}}({\bm k})$. The notation used for different quantities is the same as that defined in Table~\ref{table_1}. 
 }
	{
		\begin{tabular}{c@{\hskip 1cm} c@{\hskip 1cm} c @{\hskip 1cm}c  }
			\hline \hline
			\rule{0pt}{3ex}
			  Conductivity & Integrand &  $\mathcal{P}$ &  $\mathcal{T}$ or ${\mathcal P}{\mathcal T}$ \\ [2ex]
			\hline \hline
			\rule{0pt}{3ex}
			 $\tilde{\sigma}_{abcd,1}^{(3e);{\rm Surface}}(\bm{k})$ & $\frac{i3\tilde{\omega}}{2\tilde{\omega}^2}\sum_{nm}\frac{(\mathcal{G}_{nm}^{ad}-i\Omega_{nm}^{ad}/2)}{(\omega_{nm}-3\tilde{\omega})}\frac{\partial^2 f_{nm}}{\partial k^b \partial k^c}$ &  $\mathcal{G}$, $\Omega$ &  $\mathcal{G}$  \\ [2ex]
			 $\tilde{\sigma}_{abcd,2}^{(3e);{\rm Surface}}(\bm{k})$ & $ \frac{i3\tilde{\omega}}{\tilde{\omega}}\sum_{nm}\frac{(\mathcal{G}_{nm}^{ac}-i\Omega_{nm}^{ac}/2)}{(\omega_{nm}-3\tilde{\omega})(\omega_{nm}-2\tilde{\omega})}\frac{\partial^2 f_{nm}}{\partial k^b \partial k^d}$  &  $\mathcal{G}$, $\Omega$ &  $\mathcal{G}$  \\[2ex]
			 $\tilde{\sigma}_{abcd,3}^{(3e);{\rm Surface}}(\bm{k})$ & 	$ -\frac{i3\tilde{\omega}}{\tilde{\omega}}\sum_{nm}\frac{\omega_{nm;d}(\mathcal{G}_{nm}^{ac}-i\Omega_{nm}^{ac}/2)}{(\omega_{nm}-3\tilde{\omega})(\omega_{nm}-2\tilde{\omega})^2}\frac{\partial f_{nm}}{\partial k^b}$  &  $\mathcal{G}$, $\Omega$ & $\mathcal{G}$ \\ [2ex]
			 $\tilde{\sigma}_{abcd,4}^{(3e);{\rm Surface}}(\bm{k})$ &  $ \frac{i3\tilde{\omega}}{\tilde{\omega}}\sum_{nm}\frac{(\Gamma_{nm}^{adc}-i\tilde{\Gamma}_{nm}^{adc})}{(\omega_{nm}-3\tilde{\omega})(\omega_{nm}-2\tilde{\omega})}\frac{\partial f_{nm}}{\partial k^b }$   & $\Gamma$, $\tilde{\Gamma}$ & $\Gamma$ \\[2ex]
			 $\tilde{\sigma}_{abcd,5}^{(3e);{\rm Surface}}(\bm{k})$ & $ -i3\tilde{\omega}\sum_{nm}\frac{\omega_{nm;c}(\mathcal{G}_{nm}^{ab}-i\Omega_{nm}^{ab}/2)}{(\omega_{nm}-3\tilde{\omega})(\omega_{nm}-2\tilde{\omega})(\omega_{nm}-\tilde{\omega})^2}\frac{\partial f_{nm}}{\partial k^d}$   &  $\mathcal{G}$, $\Omega$ &  $\mathcal{G}$ \\[2ex]
			 $\tilde{\sigma}_{abcd,6}^{(3e);{\rm Surface}}(\bm{k})$ & $ i3\tilde{\omega}\sum_{nm}\frac{(\Gamma_{nm}^{acb}-i\tilde{\Gamma}_{nm}^{acb})}{(\omega_{nm}-3\tilde{\omega})(\omega_{nm}-2\tilde{\omega})(\omega_{nm}-\tilde{\omega})}\frac{\partial f_{nm}}{\partial k^d}$   & $\Gamma$, $\tilde{\Gamma}$ & $\Gamma$
             \\[2ex]	
             $\tilde{\sigma}_{abcd,7}^{(3e);{\rm Surface}}(\bm{k})$ & $ i3\tilde{\omega}\sum_{nm}\frac{(\Gamma_{nm}^{adb}-i\tilde{\Gamma}_{nm}^{adb})}{(\omega_{nm}-3\tilde{\omega})(\omega_{nm}-2\tilde{\omega})(\omega_{nm}-\tilde{\omega})}\frac{\partial f_{nm}}{\partial k^c}$   & $\Gamma$, $\tilde{\Gamma}$ & $\Gamma$
             \\[2ex]		
             $\tilde{\sigma}_{abcd,8}^{(3e);{\rm Surface}}(\bm{k})$ & $ -i3\tilde{\omega}\sum_{nm}\frac{\omega_{nm;d}(\mathcal{G}_{nm}^{ab}-i\Omega_{nm}^{ab}/2)}{(\omega_{nm}-3\tilde{\omega})(\omega_{nm}-2\tilde{\omega})(\omega_{nm}-\tilde{\omega})^2}\frac{\partial f_{nm}}{\partial k^c}$   &  $\mathcal{G}$, $\Omega$ &  $\mathcal{G}$
             \\[2ex]
             $\tilde{\sigma}_{abcd,9}^{(3e);{\rm Surface}}(\bm{k})$ & $ -i3\tilde{\omega}\sum_{nm}\frac{\omega_{nm;d}(\mathcal{G}_{nm}^{ab}-i\Omega_{nm}^{ab}/2)}{(\omega_{nm}-3\tilde{\omega})(\omega_{nm}-2\tilde{\omega})^2(\omega_{nm}-\tilde{\omega})}\frac{\partial f_{nm}}{\partial k^c}$   &  $\mathcal{G}$, $\Omega$ &  $\mathcal{G}$
             \\[2ex]
             $\tilde{\sigma}_{abcd,10}^{(3e);{\rm Surface}}(\bm{k})$ & $ -i3\tilde{\omega}\sum_{nm}\frac{(\mathcal{G}_{nm}^{ab}-i\Omega_{nm}^{ab}/2)}{(\omega_{nm}-3\tilde{\omega})(\omega_{nm}-2\tilde{\omega})(\omega_{nm}-\tilde{\omega})}\frac{\partial^2 f_{nm}}{\partial k^c \partial k^d}$   &  $\mathcal{G}$, $\Omega$ &  $\mathcal{G}$ \\[2ex]
						\hline \hline
	\end{tabular}}
	\label{table_3}
\end{table*}
\begin{table*}[t]
	\centering
 \caption{The Fermi surface contributions to the third-order intraband optical conductivity, $\sigma_{abcd}^{(3i);{\rm Surface}}(-3\omega; \omega, \omega,\omega) = e^4/\hbar^3 \sum_{\bm k} \tilde{\sigma}_{abcd}^{(3i);{\rm Surface}}({\bm k})$. The notation used for different quantities is the same as that defined in Table~\ref{table_1}. 
 }
	{
		\begin{tabular}{c@{\hskip 1cm}c@{\hskip 1cm}c@{\hskip 1cm}c  }
			\hline \hline
			\rule{0pt}{3ex}
			  Conductivity & Integrand &  $\mathcal{P}$ &  $\mathcal{T}$ or ${\mathcal P}{\mathcal T}$ \\ [2ex]
			\hline \hline
			\rule{0pt}{3ex}
			 $\tilde{\sigma}_{abcd,1}^{(3i);{\rm Surface}}(\bm{k})$ & $\frac{i}{6\tilde{\omega}^3}\sum_{n}\omega_{n;a}\frac{\partial^3 f_n}{\partial k^d \partial k^c \partial k^b}$ &  $\neq 0$ &  $\neq 0$  \\ [2ex]
			 $\tilde{\sigma}_{abcd,2}^{(3i);{\rm Surface}}(\bm{k})$ & $ \frac{1}{2\tilde{\omega}^2}\sum_{nm}\Omega_{mn}^{ad}\frac{\partial^2 f_n}{\partial k^c \partial k^b}$  &  $\Omega$  &  0  \\[2ex]
			 $\tilde{\sigma}_{abcd,3}^{(3i);{\rm Surface}}(\bm{k})$ & 	$-\frac{i}{6\tilde{\omega}^2}\sum_{nm}\frac{\omega_{nm;a}(\mathcal{G}_{nm}^{cb}-i\Omega_{nm}^{cb}/2)}{(\omega_{nm}-\tilde{\omega})}\frac{\partial f_{mn}}{\partial k^d}$ &  $\mathcal{G}$, $\Omega$ & $\mathcal{G}$ \\ [2ex]
			 $\tilde{\sigma}_{abcd,4}^{(3i);{\rm Surface}}(\bm{k})$ &  $ \frac{i}{3\tilde{\omega}} \sum_{nm}\frac{\omega_{nm;a}(\mathcal{G}_{nm}^{db}-i\Omega_{nm}^{db}/2)}{(\omega_{nm}-2\tilde{\omega})(\omega_{nm}-\tilde{\omega})} \frac{\partial f_{mn}}{\partial k^c}$  & $\cal{G}$, $\Omega$ & $\mathcal{G}$ \\[2ex]
			 $\tilde{\sigma}_{abcd,5}^{(3i);{\rm Surface}}(\bm{k})$ & $\frac{i}{\tilde{\omega}}\sum_{nm}\frac{(\Gamma_{mn}^{cad}-i\tilde{\Gamma}_{mn}^{cad})}{(\omega_{nm}-\tilde{\omega})}\frac{\partial f_{mn}}{\partial k^b}$  & $\Gamma$, $\tilde{\Gamma}$ & $\Gamma$  \\[2ex]		
						\hline \hline
	\end{tabular}}
	\label{table_4}
\end{table*}

\subsection{Third-harmonic generation}
The third-harmonic response is characterized by the third-order susceptibility tensor $\chi_{abcd}^{(3)}(-3\omega;\omega,\omega,\omega)$, or equivalently, by the conductivity tensor $\sigma_{abcd}^{(3)}(-3\omega;\omega,\omega, \omega)$. The optical susceptibility captures the polarization in the material induced by the electric field, which can be expressed as  \cite{Boyd2008_optics_book},
\be 
{\bm P}={\bm P}_0 +\chi^{(1)}{\bm E}+\chi^{(2)}{\bm E}^2+ \chi^{(3)}{\bm E}^3+\cdots~.
\ee 
Here, ${\bm P}_0$ is the electric polarization in the absence of an external electric field, $\chi^{(1)}$ is the linear susceptibility, and $\chi^{(2)}$, $\chi^{(3)}, \ldots$ are nonlinear susceptibilities. These nonlinear terms of ${\bm P}$ produce responses at new frequencies via harmonic generation and frequency mixing. For instance, a second-harmonic response at $2\omega$ arises from an input at $\omega$, while THG yields a response at $3\omega$. The polarization operator is given by $\hat{\bm P}=e\hat{\bm r}$, and in terms of density matrix, we have 
\be
{\bm P} = {\rm Tr}[e\hat{\bm r}\rho] =  e{\rm Tr}[\hat{\bm r}_i\rho] + e{\rm Tr}[\hat{\bm r}_e\rho] ~,
\ee
where we have separated the intraband and interband contributions. Accordingly, the third-order susceptibility, defined as 
\be
P^{(3)}_a=\sum\chi_{abcd}^{(3)}(-3\omega;\omega,\omega, \omega)E_b E_c E_d e^{-i3\omega t}~,
\ee
naturally separates into interband ($\chi^{(3e)}$) and intraband ($\chi^{(3i)}$) components. The interband polarization can be written as 
\be \label{chi_3e}
{P}^{(3e)}_a=e\sum_{nm}\mathcal{R}_{mn}^a\rho_{nm}^{(3)}~.
\ee 
Here, $\mathcal{R}_{mn}^a$ is the $a$-th component of the band-resolved Berry connection. The intraband contribution, however, cannot be directly calculated using the position matrix elements due to their singular behavior [see Eq.\eqref{position_matrix}].  Instead, we derive it from the polarization current ${\bm J}=\frac{d{\bm P}}{dt}$. The intraband contribution is obtained from the total current after subtracting the interband part~\cite{Fregoso2019_bulk}, 
\bea \label{J_3i}
J^{(3i)}_a&=&e\sum_{nm}\left[\omega_{n;a}\rho_{nn}^{(3)} - \frac{e}{\hbar}\left({\bm E}\times {\bm \Omega}_n\right)^a\rho_{nn}^{(2)}\right.\nn\\ &-& \left.\frac{e}{\hbar}{\bm E}\cdot \bm{\mathcal{R}}_{mn;a}\rho_{nm}^{(2)}\right]~.
\eea
The susceptibility is related to the conductivity via the relation, 
\be \label{chi_sigma_3rd_order}
\chi_{abcd}^{(3)}(-3\omega;\omega,\omega, \omega)=\frac{\sigma^{(3)}_{abcd}(-3\omega;\omega,\omega, \omega)}{-i3\omega}~.
\ee 
We present the obtained Fermi sea contributions to the interband and intraband responses in Table~\ref{table_1} and Table~\ref{table_2}, respectively. Likewise, the Fermi surface contributions to the interband and intraband responses are presented in Table~\ref{table_3} and Table~\ref{table_4}. {Note that the conductivities listed in Tables~\ref{table_1}–\ref{table_4} are not field-symmetrized. The symmetrized conductivity is obtained via  
\be
\bar{\sigma}_{abcd}^{(3)}=\frac{1}{6}\left[\sigma_{abcd}^{(3)}+\sigma_{abdc}^{(3)}+\sigma_{adbc}^{(3)}+\sigma_{adcb}^{(3)}+\sigma_{acdb}^{(3)}+\sigma_{acbd}^{(3)}\right]~,
\ee
where the frequency arguments $(-3\omega,\omega,\omega,\omega)$ are implied for each term.}
The geometry of the Bloch states plays a crucial role in shaping these optical responses. Our analysis shows that THG is governed by five key quantum geometric quantities: quantum metric ($\mathcal{G}_{nm}^{ab}$), Berry curvature ($\Omega_{nm}^{ab}$), metric connection ($\Gamma_{nm}^{abc}$), symplectic connection ($\tilde{\Gamma}_{nm}^{abc}$), and the second-order connection $(D_{nm}^{abcd})$~\cite{Ahn2021_Riemannian}. Notably, the first four quantities represent the real and imaginary components of two complex geometric tensors: the quantum geometric tensor ($\mathcal{Q}_{nm}^{ab}$) \cite{Bhalla2022_resonant, Bhalla2023} and the geometric connection ($\mathcal{C}_{nm}^{abc}$) \cite{Ahn2020_low, Watanabe2021_chiral}, defined as
\bea \label{QGT_eq}
\mathcal{Q}_{nm}^{ab} &=& \mathcal{R}_{mn}^a \mathcal{R}_{nm}^b = \mathcal{G}_{nm}^{ab} - \frac{i}{2} \Omega_{nm}^{ab}~,
\\
\label{QGC_eq}
\mathcal{C}_{nm}^{abc} &=& \mathcal{R}_{mn}^a \mathcal{R}_{nm;b}^c = \Gamma_{nm}^{abc} - i\tilde{\Gamma}_{nm}^{abc} ~.
\eea
Here, $\mathcal{R}_{nm;b}^c=\partial_b \mathcal{R}_{nm}^c-i(\mathcal{R}_{nn}^b-\mathcal{R}_{mm}^b)\mathcal{R}_{nm}^c$ is the covariant derivative of $\mathcal{R}_{nm}^c$ with respect to Bloch momentum $k_b$. Additionally, the second-order connection is defined as, $D_{nm}^{abcd}=\mathcal{R}_{nm}^a(\mathcal{R}_{nm}^b)_{;cd}$, with $(\mathcal{R}_{nm}^b)_{;cd}$ being the second order covariant derivative of $\mathcal{R}_{nm}^b$. Together, these results establish a comprehensive theoretical framework for THG in terms of quantum band geometric quantities. We now focus on the polarization dependence of the THG response. 

\begin{figure}[t!] 
    \includegraphics[width=\linewidth]{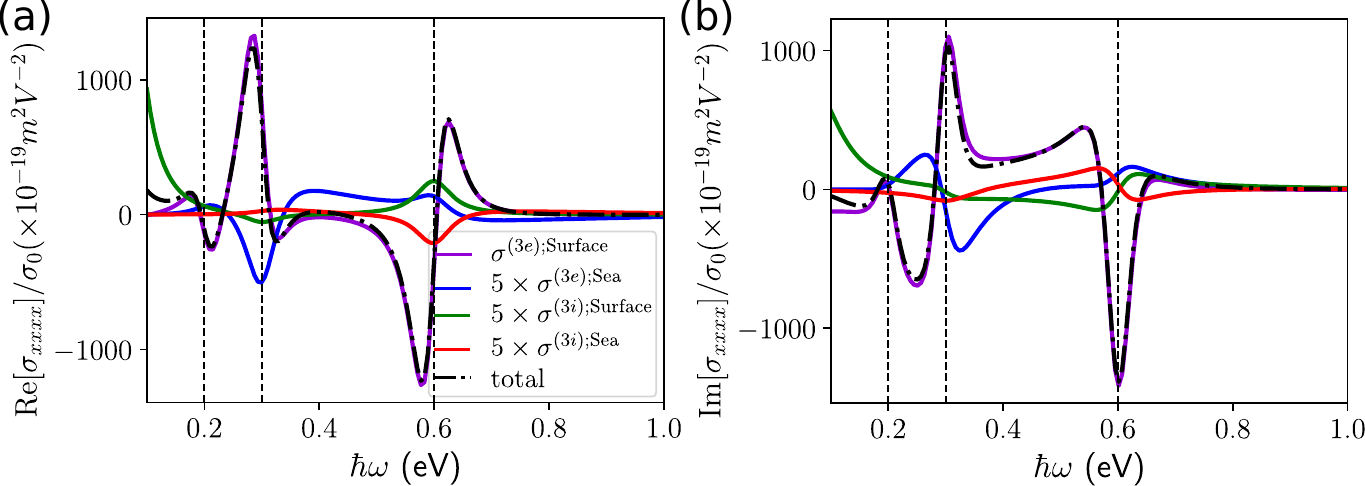}
    \caption{ 
     (a)The real and (b) imaginary part of the third-harmonic conductivity as a function of frequency in pristine graphene. Resonances appear at $\hbar\omega=\frac{2}{3}\mu$, $\hbar\omega=\mu$ and $\hbar\omega=2\mu$, as indicated by the dotted vertical lines. Here, $\sigma_0=e^2/4\hbar$. We have used $a=1.42$ ${\rm \AA}$, $t=2.7$ eV, $\mu=0.3$ eV, $\tau=2\times 10^{-14}$ s, and temperature $T=30$ K. 
\label{Fig1}}
\end{figure}

\section{THG under different polarization angles}\label{polar}
Let's consider a linearly polarized (LP) incident light beam. In an experimental scenario, the polarization angle dependence of third-harmonic response along and perpendicular to the applied electric field can be probed.  

The third-harmonic current is given by
\bea \label{Ja}
J_a(3\omega) &=& \sum_{bcd} \sigma_{abcd}(-3\omega;\omega,\omega,\omega) 
E_b(\omega) E_c(\omega) E_d(\omega) e^{-i3\omega t} 
\nn\\&+& {\rm c.c.}~
\eea
To capture the polarization dependence of THG, we consider a normally incident field, and derive an effective conductivity that reflects phase-sensitive responses. Consider an incident electric field of the form
\be
{\bm E}(t) = |E| (\cos\theta,\, \sin\theta,\,0)\, e^{-i \omega t}+{\rm c.c.}~,
\ee
where $\theta$ denotes the angle between the field and the crystallographic $\hat{x}$ axis.  

Since $\sigma_{abcd}$ is symmetric under permutation of the last three indices $(b,c,d)$, Eq.~\eqref{Ja} allows us to express the  effective THG response in terms of a complex effective third-order conductivity, $\sigma_a^{\rm eff}(\theta) = \sigma_a^{\prime}(\theta) + i \sigma_a^{\prime \prime}(\theta)$. These terms arise from the THG tensor elements weighted by angular factors isolating in-phase and out-of-phase contributions. The resulting current along the $a$-axis can be expressed as 
\be
J_a(3 \omega, \theta,t) 
= |E|^3 \Big[ \sigma_a^{\prime}(\theta)\cos(3\omega t) 
+ \sigma_a^{\prime \prime}(\theta)\sin(3\omega t)\Big]~,
\ee
Here, the effective conductivity is given by, $\sigma_a^{\rm eff}=\sigma_{axxx}\cos^3\theta+3\sigma_{axxy}\cos^2\theta\sin\theta+3\sigma_{axyy}\cos\theta\sin^2\theta+\sigma_{ayyy}\sin^3\theta$. Its real and imaginary components can be easily obtained using the corresponding real and imaginary parts of the THG response tensors. 
Using these effective in-plane responses, we can further define THG conductivity components parallel and perpendicular to the applied electric field. We obtain, 
\bea
\sigma^{(3)}_{\parallel}(\theta) &=& 
\sigma_x^{\rm eff}(\theta)\cos\theta + \sigma_y^{\rm eff}(\theta)\sin\theta~, \\
\sigma^{(3)}_{\perp}(\theta) &=& 
-\sigma_x^{\rm eff}(\theta)\sin\theta + \sigma_y^{\rm eff}(\theta)\cos\theta~.
\eea
In terms of the original THG responses, we obtain, 
\begin{widetext}
\bea\label{Eq_pol1}
\sigma^{(3)}_{\parallel}(\theta)
&=& \frac{3}{2}(\sigma_{xxxx} + \sigma_{xxyy})
+ \frac{1}{2}(\sigma_{xxxx} - 3\sigma_{xxyy})\cos(4\theta)
+ (3\sigma_{xxxy} + \sigma_{yxxx})\sin(2\theta)~, \\
\sigma^{(3)}_{\perp}(\theta)
&=& \frac{1}{2}(-\sigma_{xxxx} + 3\sigma_{xxyy})\sin(4\theta)
+ 2\sigma_{yxxx}\cos(2\theta)~.\label{Eq_pol2}
\eea
\end{widetext}

This decomposition clearly separates the angular structure of the THG response. 
The parallel component, $\sigma^{(3)}_{\parallel}$, contains a constant term along with 
$\cos(4\theta)$ and $\sin(2\theta)$ modulations. In contrast,  the perpendicular component $\sigma^{(3)}_{\perp}$ is governed by $\sin(4\theta)$ and $\cos(2\theta)$ terms. 
The real and imaginary components of $\sigma^{(3)}_{\parallel}$ and $\sigma^{(3)}_{\perp}$ will capture the in-phase and out-of-phase responses, respectively.  
The relative weight of these harmonics is set by the independent conductivity tensor elements, which depend on the crystalline symmetries. This angular dependence, which is absent in DC third-order response studies, enables the tuning of THG in different materials and captures the role of crystalline symmetries in THG response patterns. For example, Fig.~\ref{fig5} in Sec.~\ref{model_sec} illustrates these modulations for RuO$_2$ in its altermagnetic phase. 

\begin{figure}[t!] 
    \includegraphics[width=\linewidth]{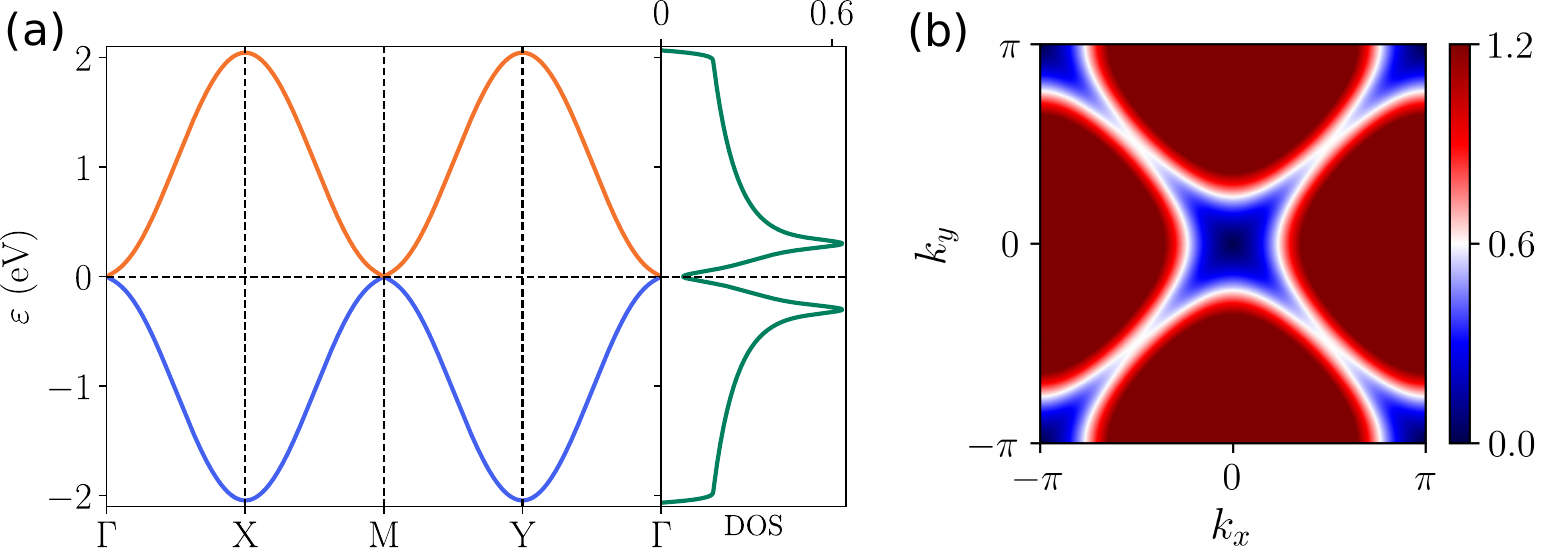}
    \caption{(a) The electronic band structure of a \(d_{x^2 - y^2}\)-wave altermagnet along the high-symmetry path $\Gamma$–X–M–Y-$\Gamma$ in the Brillouin zone. We used $J=1.0 ~{\rm eV}$ and $\lambda=0.3~{\rm eV}$. The corresponding density of states on the right panel, has van hove singularities at energies $\sim \pm \lambda = \pm .3$ eV. (b) The band splitting, $\Delta \varepsilon_k$ between the two  bands over the Brillouin zone. The white lines mark contour (energy $ = 2 \lambda$) and the states related to the van Hove singularities seen in (a).
\label{fig2}}
\end{figure}

\section{THG in graphene}\label{THG_graphene}
As a check of our calculations, we calculate the THG in pristine monolayer graphene and compare our results with the earlier works of Cheng et al.~\cite{cheng_NJP2014_third} and Mikhailov~\cite{Mikhailov2014_graphene}. The tight-binding Hamiltonian for graphene, considering only nearest-neighbor hopping, is a $2 \times 2$ matrix with elements $\mathcal{H}_{11} = \mathcal{H}_{22} = 0$, and $\mathcal{H}_{12} = \mathcal{H}^*_{21} = t f(\bm{k})$. Here, $t$ is the hopping parameter and $f(\bm{k})=e^{-ik_x a}\left[1+2 e^{i3k_x a/2} \cos \left(k_y a\sqrt{3}/2\right)\right]$, with $a$ being the lattice constant. Cheng et al. reported resonances at $\hbar\omega = 2\mu/3$ and $\hbar\omega = \mu$ in the THG response of graphene, but found no resonance enhancement at $\hbar\omega = 2\mu$. Conversely, Mikhailov observed the $\hbar\omega = 2\mu$ resonance but did not capture the lower-energy resonances at $2\mu/3$ and $\mu$. Our numerical calculations for pristine graphene, presented in Fig.~\ref{Fig1}, reveals all three resonance peaks at $\hbar\omega = \frac{2}{3}\mu$, $\hbar\omega = \mu$, and $\hbar\omega = 2\mu$. These correspond to resonant interband transitions involving the absorption of three, two, and one photon(s), respectively. Thus, our results build upon the  results of Refs.~\cite{cheng_NJP2014_third, Mikhailov2014_graphene} by  capturing the complete set of resonances of THG in graphene within a single framework. Having validated our theory on graphene, we next apply it to investigate band geometry-driven THG in an emerging class of materials, altermagnets. 

\begin{figure}[b!]
    \includegraphics[width=\linewidth]{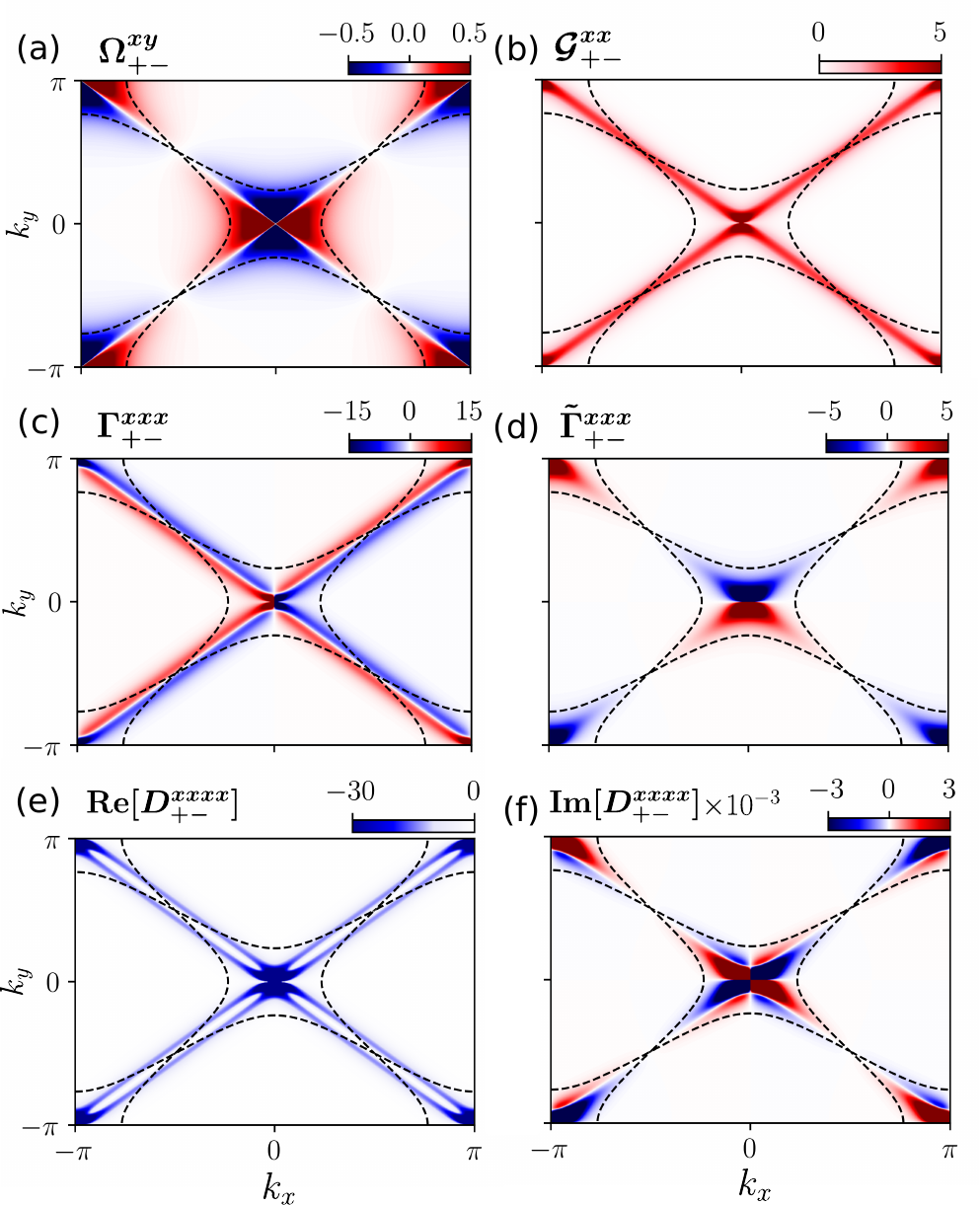}
    \caption{Band geometric quantities. The momentum space distribution of (a) the Berry curvature, (b) the quantum metric, (c) metric connection, (d) symplectic connection, (e) real and (f) imaginary part of second-order connection. The dashed line (black) represents the Fermi surface contour at $\mu=\lambda$. 
    \label{fig3}}
\end{figure}

\section{THG in altermagnet\label{model_sec}}
To demonstrate THG in altermagnets, we consider even parity altermagnets in which the dominant nonlinear optical response is third order. In general altermagnets or magnets with 
\(C_n\mathcal{T}\) symmetry, the dominant nonlinear optical process is third order. In this section, we demonstrate THG in a quasi-2D \(d_{x^2 - y^2}\)-wave altermagnet, using a minimal two-band model that captures its essential physics~\cite{libor_PRX2022_beyond, fang_PRL2024_quan}. The model is defined on a square lattice with magnetic atoms at \(A = (\frac{1}{2}, 0)\) and \(B = (0, \frac{1}{2})\), forming two interpenetrating sublattices. In the altermagnetic phase, the \(A\) and \(B\) sites host staggered magnetic moments with alternating up- and down-spin orientations. This staggered arrangement breaks time-reversal symmetry while preserving inversion symmetry. The effective Hamiltonian is given by~\cite{fang_PRL2024_quan}, 
\begin{eqnarray} \label{Ham}
\mathcal{H}&=&J(\cos k_x-\cos k_y)\sigma_z + \lambda\left[\sin\left(\frac{k_x+k_y}{2}\right)\sigma_x \right. \nonumber\\
&+& \left. \sin\left(\frac{k_y-k_x}{2}\right)\sigma_y \right]~.
\end{eqnarray}
Here, \(J\) is the altermagnetic order parameter linked to magnetic anisotropy, and \(\lambda\) is the spin-orbit coupling strength. The Hamiltonian respects a combined fourfold rotational and time-reversal symmetry (\(C_4\mathcal{T}\)) and belongs to the magnetic point group \(4'/mm'm\). Inversion symmetry ensures that all even-order optical responses vanish, making third-order effects such as THG the leading nonlinear response in the system.  

\begin{figure}[t!] 
    \includegraphics[width=\linewidth]{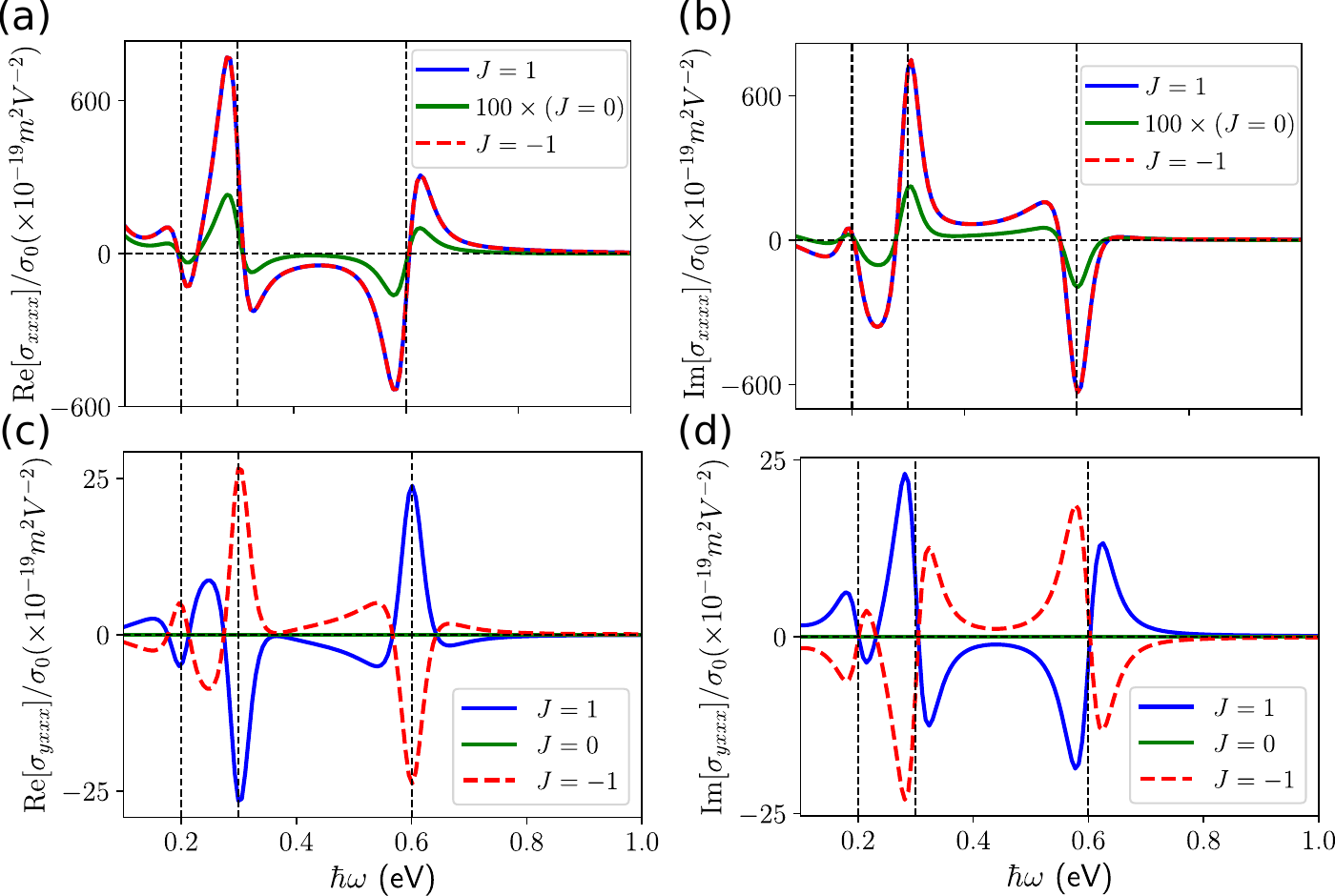}
    \caption{ 
     (a,b) The real and imaginary part of the longitudinal third-harmonic conductivity $\sigma_{xxxx}$
, and (c,d) the real and imaginary parts of the transverse third-harmonic conductivity $\sigma_{yxxx}$
, in a two-dimensional altermagnet as functions of frequency. Resonances appear at $\hbar\omega=\frac{2}{3}\mu$, $\hbar\omega=\mu$ and $\hbar\omega=2\mu$, as indicated by the dotted vertical lines. Here, $\sigma_0=e^2/4\hbar$. We have used $J=1.0~{\rm eV}$, $\lambda=\mu=0.3$ eV, $\tau=2\times 10^{-14}$ s, and temperature $T=30$ K. 
\label{fig4}}
\end{figure}

The energy dispersion for the altermagnet is given by 
\be \varepsilon_k=\pm\sqrt{\lambda^2(1-\cos k_x \cos k_y)+J^2(\cos k_x-\cos k_y)^2}~.
\ee 
Here, $+(-)$ denotes the conduction (valence) band.  We present the band structure along with the density of states (DOS) in Fig.~\ref{fig2}(a). The presence of the $C_4 {\mathcal T}$ symmetry can be checked from the dispersion as $\varepsilon (-k_y, k_x ) = \varepsilon (k_x, k_y )$. 
Fig.~\ref{fig2}(b) shows the energy dispersion, $\Delta \varepsilon_k = \varepsilon_{k}^{\rm conduction} - \varepsilon_{k}^{\rm valance}$ in the momentum space, with the white lines marking the contour of energy at which the VHS appears. The blue region near the band edges close to $0$ energy, captures the region where the band geometry is significant. Figure~\ref{fig3}(a-b) shows the momentum-resolved Berry curvature and quantum metric of the altermagnetic model. The metric connection, symplectic connection, and real and imaginary parts of the second-order connection are displayed in Fig.~\ref{fig3}(c–f). These geometric quantities peak sharply near the Dirac points \(\Gamma = (0, 0)\), \(M = (\pi, \pi)\), and along the anti-crossing lines \(k_x = \pm k_y\), indicating regions of strong band overlap.

In Fig.~\ref{fig4}(a–b), we present the frequency dependence of the real and imaginary parts of the longitudinal third-order conductivity \(\sigma_{xxxx}\) for different \(J\) values. Figures~\ref{fig4}(c–d) show the corresponding transverse component \(\sigma_{yxxx}\). The longitudinal response is even under \(J \to -J\) and remains finite as \(J \to 0\), indicating geometric contributions that persist without magnetic order. In contrast, the transverse conductivity is odd under \(J \to -J\) and vanishes in the \(J \to 0\) limit, reflecting its direct dependence on time-reversal symmetry breaking. 

Remarkably, we find that all third-order conductivity components display sharp resonances at \(\hbar\omega = \frac{2}{3}\mu\), \(\hbar\omega = \mu\), and \(\hbar\omega = 2\mu\), marked by vertical dashed lines in Fig.~\ref{fig4}. These features correspond to resonant interband transitions involving the absorption of three, two, and one photon(s) across the Pauli blocked region of the electronic bands, respectively. The tunability of these resonances through doping or gating provides clear experimental signatures for their detection. 

\begin{figure}[t!]
    \includegraphics[width=\linewidth]{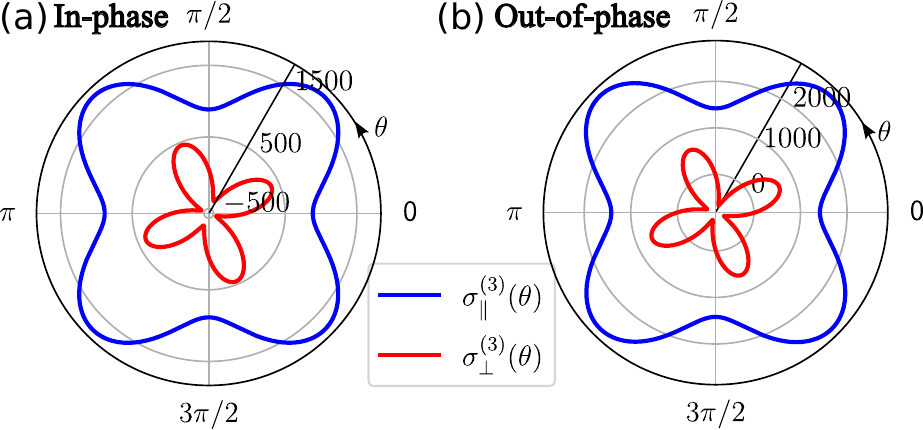}
    \caption{Polarization-angle dependence of the third-harmonic generation (THG) conductivity, along ($\sigma_{\parallel}^3$) the applied electric field and perpendicular ($\sigma_{\perp}^3$) to it, for a $d_{x^2-y^2}$-wave altermagnet. We have chosen the optical frequency $\hbar\omega=\mu$. (a) in-phase THG conductivity and (b) out-of-phase THG conductivity. The parameters are the same as in Fig.~\ref{fig4}. This highlights the tunability of the THG responses by varying the polarization angle. 
\label{fig5}}
\end{figure}

We now analyze the polarization-angle dependence of the effective THG conductivities 
for LP light, as defined in Eqs.~\eqref{Eq_pol1}--\eqref{Eq_pol2}. 
Figure~\ref{fig5}(a) shows the components of the in-phase conductivity projected parallel 
and perpendicular to the incident electric field, while Fig.~\ref{fig5}(b) presents the corresponding 
out-of-phase components.
For the $d_{x^2-y^2}$-wave altermagnet, the THG tensor elements  $\sigma_{xxxy}$ and $\sigma_{yxxx}$ are  
much smaller than $\sigma_{xxxx}$ and $\sigma_{xxyy}$, leading to a dominant $\cos(4\theta)$ and $\sin(4\theta)$ modulation in the angular dependence. This illustrates the polarization angle dependence and tunability of the THG responses in RuO$_2$.  Beyond this, understanding the symmetry constraints of THG responses is crucial for material discovery. We focus on this aspect in the next section. 

\begin{table*}
\renewcommand{\arraystretch}{1.2}
\setlength{\tabcolsep}{4pt}
\caption{ List of the magnetic point groups demonstrating finite third-harmonic conductivity. Here we have limited our attention in the planar setup, where the response is measured in the same plane the external electric field is applied. The ${\mathcal T}$-odd response vanishes in the nonmagnetic materials, while the ${\mathcal T}$-even response is finite in both magnetic and nonmagnetic materials. 
\label{table_thg_mpg_2d_total}
}
    \begin{tabular}{lll}
    \hline\hline
    $\bar{\sigma}_{a;bcd}$ & ${\mathcal T}$-\textbf{even} & ${\mathcal T}$-\textbf{odd}
    \\
    \hline\hline
    & \multirow{5}{*}{~~~~~~~~~~~~~~~~~~~~~~~~~ -} & \textbf{Grey mpgs}, $-1^\prime$, $2^\prime/m$, $2/m^\prime$, $m^\prime mm$,
    \\
    \textit{(R1) All components prohibited} &  & $m^\prime m^\prime m^\prime$, $4/m^\prime$, $4^\prime/m^\prime$, $4/m^\prime mm$, $4^\prime/m^\prime m^\prime m$,
    \\
    &  & $4/m^\prime m^\prime m^\prime$, $-3^\prime$, $-3^\prime m$, $-3^\prime m^\prime$, $6^\prime/m$, $6/m^\prime$,
    \\
    &  & $6/m^\prime mm$, $6^\prime/mmm^\prime$, $6/m^\prime m^\prime m^\prime$, $m^\prime-3^\prime$,
    \\
    &  & $m^\prime-3^\prime m$, $m^\prime-3^\prime m^\prime$
    \\
    \hline
    \textit{(R2) No planar response } &  & $m^\prime m2^\prime$, $6^\prime$, $-6^\prime$, $6^\prime/m^\prime$, $6^\prime22^\prime$, $6^\prime mm^\prime$,
    \\
    &  & $-6^\prime m^\prime2$, $-6^\prime m2^\prime$, $6^\prime/m^\prime mm^\prime$
    \\
    \hline
    \textit{(R3) All components allowed} & $1$, $1^\prime$, $-1$, $-11^\prime$, $-1^\prime$ & $1$, $-1$
    \\
    \hline
    \multirow{4}{*}{\textit{(R4)} $xxxx$, $yyyy$, $xxyy$, $yyxx$ } & $2$, $21^\prime$, $2^\prime$, $m$, $m1^\prime$, $m^\prime$, $2/m$, $2/m1^\prime$, & \multirow{4}{*}{$2$, $m$, $2/m$, $222$, $mm2$, $mmm$ }
    \\
    & $2^\prime/m$, $2/m^\prime$, $2^\prime/m^\prime$, $222$, $2221^\prime$, $2^\prime2^\prime2$, & 
    \\
    & $mm2$, $mm21^\prime$, $m^\prime m2^\prime$, $m^\prime m^\prime2$, $mmm$, &
    \\
    & $mmm1^\prime$, $m^\prime mm$, $m^\prime m^\prime m$, $m^\prime m^\prime m^\prime$ & 
    \\
    \hline
    \textit{(R5)} $xxxy$, $yyyx$, $xyyy$, $yxxx$ & ~~~~~~~~~~~~~~~~~~~~~~~~~ - & $2^\prime$, $m^\prime$, $2^\prime/m^\prime$, $2^\prime2^\prime2$, $m^\prime m^\prime2$, $m^\prime m^\prime m$
    \\
    \hline
    \textit{(R6)} $yyyy=xxxx$, $yyyx=-xxxy$, & $4$, $41^\prime$, $4^\prime$, $-4$, $-41^\prime$, $-4^\prime$, $4/m$, & \multirow{2}{*}{$4$, $-4$, $4/m$ }
    \\
    ~~~~~~ $yxxx=-xyyy$, $yyxx=xxyy$ & $4/m1^\prime$, $4^\prime/m$, $4/m^\prime$, $4^\prime/m^\prime$ & 
    \\
    \hline
    \textit{(R7)} $yxxx=xyyy$, $yyyx=xxxy$ & ~~~~~~~~~~~~~~~~~~~~~~~~~ - & $4^\prime m^\prime m$, $-4^\prime2^\prime m$, $4^\prime/mm^\prime m$
    \\
    \hline
    \multirow{7}{*}{\textit{(R8)} $yyyy=xxxx$, $yyxx=xxyy$ } & $422$, $4221^\prime$, $4^\prime22^\prime$, $42^\prime2^\prime$, $4mm$, $4mm1^\prime$, & 
    \\
    & $4^\prime m^\prime m$, $4m^\prime m^\prime$, $-42m$, $-42m1^\prime$, $-4^\prime2^\prime m$, & 
    \\
    & $-4^\prime2m^\prime$, $-42^\prime m^\prime$, $4/mmm$, $4/mmm1^\prime$, & $422$, $4mm$, $-42m$, $4/mmm$, $432$, $-43m$,
    \\
    & $4/m^\prime mm$, $4^\prime/mm^\prime m$, $4^\prime/m^\prime m^\prime m$, & $m-3m$
    \\
    & $4/mm^\prime m^\prime$, $4/m^\prime m^\prime m^\prime$, $432$, $4321^\prime$, $4^\prime32^\prime$, & 
    \\
    & $-43m$, $-43m1^\prime$, $-4^\prime3m^\prime$, $m-3m$, & 
    \\
    & $m-3m1^\prime$, $m^\prime-3^\prime m$, $m-3m^\prime$, $m^\prime-3^\prime m^\prime$ & 
    \\
    \hline
    \textit{(R9)} $yyyy=-xxxx$, $yyxx=-xxyy$ ~ & ~~~~~~~~~~~~~~~~~~~~~~~~~ - & $4^\prime22^\prime$, $-4^\prime2m^\prime$ 
    \\
    \hline
    \textit{(R10)} $yxxx=-xyyy$, $yyyx=-xxxy$ & ~~~~~~~~~~~~~~~~~~~~~~~~~ - & $42^\prime2^\prime$, $4m^\prime m^\prime$, $-42^\prime m^\prime$, $4/mm^\prime m^\prime$
    \\
    \hline
    \textit{(R11)} $yyxx=-xxyy$ & ~~~~~~~~~~~~~~~~~~~~~~~~~ - & $4^\prime32^\prime$, $-4^\prime3m^\prime$, $m-3m^\prime$
    \\
    \hline
    \textit{(R12)} $yyyy=-xxxx$, $yyyx=xxxy$, & \multirow{2}{*}{~~~~~~~~~~~~~~~~~~~~~~~~~ -} & \multirow{2}{*}{ $4^\prime$, $-4^\prime$, $4^\prime/m$ }
    \\
    ~~~~~~~~ $yyxx=-xxyy$, $yxxx=xyyy$ &  & 
    \\
    \hline
    \textit{(R13)} $yyxx = xxyy = \tfrac{1}{3}yyyy = \tfrac{1}{3}xxxx$, & $3$, $31^\prime$, $-3$, $-31^\prime$, $-3^\prime$, $6$, $61^\prime$, $6^\prime$, $-6$, $-61^\prime$, & \multirow{2}{*}{ $3$, $-3$, $6$, $-6$, $6/m$ }
    \\
    ~~~~~~~~ $yxxx=3yyyx=-3xxxy=-xyyy$ & $-6^\prime$, $6/m$, $6/m1^\prime$, $6^\prime/m$, $6/m^\prime$, $6^\prime/m^\prime$ &   
    \\
    \hline
    \multirow{6}{*}{\textit{(R14)} $yyxx=xxyy=\tfrac{1}{3}yyyy=\tfrac{1}{3}xxxx$} & $32$, $321^\prime$, $32^\prime$, $3m$, $3m1^\prime$, $3m^\prime$, $-3m$, & 
    \\
    & $-3m1^\prime$, $-3^\prime m$, $-3^\prime m^\prime$, $-3m^\prime$, $622$, $6221^\prime$, &
    \\
    & $6^\prime22^\prime$, $62^\prime2^\prime$, $6mm$, $6mm1^\prime$, $6^\prime mm^\prime$, & $32$, $3m$, $-3m$, $622$, $6mm$, $-6m2$, $6/mmm$
    \\
    & $6m^\prime m^\prime$, $-6m2$, $-6m21^\prime$, $-6^\prime m^\prime2$, $-6^\prime m2^\prime$, & 
    \\
    & $-6m^\prime 2^\prime$, $6/mmm$, $6/mmm1^\prime$, $6/m^\prime mm$, &
    \\
    & $6^\prime/mmm^\prime$, $6^\prime/m^\prime mm^\prime$, $6/mm^\prime m^\prime$, $6/m^\prime m^\prime m^\prime$ &
    \\
    \hline
    \multirow{2}{*}{\textit{(R15)} $yxxx=-xyyy=-3xxxy=3yyyx$} & \multirow{2}{*}{~~~~~~~~~~~~~~~~~~~~~~~~~ -} & $32^\prime$, $3m^\prime$, $-3m^\prime$, $62^\prime2^\prime$, $6m^\prime m^\prime$, $-6m^\prime 2^\prime$,
    \\
    &  & $6/mm^\prime m^\prime$
    \\
    \hline
    \textit{(R16)} $yyyy=xxxx$, $yyxx$, $xxyy$ & $23$, $231^\prime$, $m-3$, $m-31^\prime$, $m^\prime-3^\prime$ & $23$, $m-3$
    \\
    \hline\hline
    \end{tabular}
\end{table*}
\section{Symmetry analysis of third-harmonic response\label{sym_sec}}

Having demonstrated THG in an altermagnet, we now focus on the symmetry properties of the THG responses. This helps identify materials which support specific THG responses, for experimental demonstration and potential applications. The existing crystalline symmetry analysis of THG is restricted to the 32 nonmagnetic gray point groups~\cite{Popov1995_suscept_book}. We build on this to provide a complete and more comprehensive symmetry classification of third-harmonic conductivity in all 122 magnetic point groups, which includes magnetic materials.  

Using the MTENSOR program~\cite{Gallego2019_acta_sym} of the Bilbao Crystallographic Server, we determine the symmetry-imposed restrictions on the third-harmonic conductivity tensor $\sigma^{(3)}(-3\omega;\omega,\omega,\omega)$ for all 122 magnetic point groups (MPGs). Each MPG contains a set of symmetry operations whose combined action dictates the allowed response tensors. MTENSOR requires the Jahn symbol of a tensor as input, and hence we first determine the Jahn symbol of the third-harmonic conductivity from its fundamental physical characteristics.
The third-harmonic conductivity $\sigma^{(3)}(-3\omega;\omega,\omega,\omega)$ is a fourth-rank polar tensor. 
Because all applied fields have the same frequency ($\omega$), the tensor is symmetric in the spatial field indices. Further, the conductivity can be decomposed into a $\mathcal{T}$-even part, which remains unchanged under $\mathcal{T}$, and a $\mathcal{T}$-odd part, which changes sign. Using the Jahn symbol notation, the $\mathcal{T}$-even and $\mathcal{T}$-odd parts correspond to $V[V3]$ and $aV[V3]$, respectively, which serve as inputs to MTENSOR for predicting the symmetry restrictions. We summarize the results of our symmetry analysis in Table~\ref{table_thg_mpg_2d_total}.

Within each MPG, every symmetry element imposes specific constraints on the THG conductivity tensor. By Neumann’s principle, a system’s physical tensors must be invariant under all its symmetry operations. 
For a system with a combined symmetry $R\mathcal{T}$, the field-symmetrized third-harmonic conductivity $\bar{\sigma}^{(3)}(-3\omega;\omega,\omega,\omega)$ transforms as
\be \label{neumann}
\bar{\sigma}_{abcd}^{(3)}=\eta_{\mathcal T}R_{aa^\prime}R_{bb^\prime}R_{cc^\prime}R_{dd^\prime}\bar{\sigma}_{a^\prime b^\prime c^\prime d^\prime}^{(3)}~.
\ee
Here, $R$ denotes a spatial symmetry operation, such as inversion (${\mathcal P}$), mirror ($\sigma_{i=x,y,z}$) operations, proper rotations ($C_n$) and improper rotations ( $S_n$). The factor $\eta_{\mathcal{T}}$ equals $+1$ for the $\mathcal{T}$-even conductivity and $-1$ for the $\mathcal{T}$-odd conductivity. Applying Eq.~\eqref{neumann} for the spatial symmetries, we find that the ${\mathcal T}$-odd part of the third-harmonic conductivity vanishes for ${\mathcal T}$, ${\mathcal PT}$, $C_3^{(x,z)}{\mathcal T}$ and $S_6^{(x,z)}{\mathcal T}$-symmetric systems, while there is no such restriction for the ${\mathcal T}$-even part.

Some significant trends from Table~\ref{table_thg_mpg_2d_total} are as follows. Among the 122 MPGs, the ${\mathcal T}$-odd response is prohibited in 53 MPGs (row R1 in the Table), including the 32 grey point groups where time-reversal is a symmetry element. Nine MPGs in R2 prohibit any in-plane response when the applied electric field is also in the same plane. Row R3 reinforces that third-harmonic current is finite in centrosymmetric materials, whereas the ${\mathcal PT}$-symmetric systems strictly allow only a ${\mathcal T}$-even response, prohibiting the ${\mathcal T}$-odd components. The remaining rows (R4-R16) list the nonzero tensor components allowed by different MPGs and the symmetry relations between them.  

These symmetry insights, combined with our geometric expressions, provide a foundation for THG engineering. Yet a few practical aspects, including computational choices and disorder, need further discussion. 

\section{Discussion} \label{discussion}

Our calculations employ the length gauge, which is formally equivalent to the velocity gauge in a complete band basis~\cite{ventura_PRB2017_gauge}. However, practical implementations often employ a truncated set of bands, and then the results for nonlinear responses obtained using different gauge choices can be different. In practical implementations, the length gauge offers a few advantages. It directly couples to the electric field, avoiding spurious low-frequency divergences common in the velocity gauge~\cite{Morimoto2016a}, and ensures a balanced treatment of intra- and inter-band contributions without relying on sum-rule cancellations. This approach also provides faster numerical convergence and more accurate spectra~\cite{taghizadeh_PRB2017_lin}, making it well-suited for THG studies based on density functional theory based first principle calculations. 

Our current framework incorporates the effects of symmetric disorder scattering through a phenomenological relaxation time $\tau$. Going beyond the constant relaxation time approximation, a more realistic modeling of $\tau$ for specific kinds of static and dynamic disorder can be incorporated in the current framework. Including detailed scattering modeling will lead to energy and momentum dependence of the scattering timescale, which can be included in the integration kernel of the responses.  Moreover, the third-order responses can have additional contributions arising from asymmetric scattering mechanisms like side-jump and skew scattering processes~\cite{Yu2025_the,Barman2025_disorder,Du2019,CXiao2019,Du2021}. These asymmetric scattering mechanisms are known to play an important role in second-order nonlinear transport responses~\cite{Datta2024, ahmed_small2025}, and represent a promising direction for further exploration. 

The predicted third-harmonic susceptibility can be easily probed in experiments. THG is usually quantified using the ratio of output power $P_{3\omega}$ to the incident power $P_\omega$ at frequency $\omega$ ~\cite{Soavi2018_broadbandthg,Jiang2018_gate,Zhu2025_bp_mid_infra}. The third-harmonic susceptibility $\chi_{3\omega}$ is given by, 
\be
P_{3\omega} = F(\omega, \alpha, d, n_\omega, n_{3\omega}, \Delta k, f_{\rm rep}, \tau_p) |\chi_{3\omega}|^2 P_\omega^3~.
\ee
Here, $F$ accounts for parameters like the absorption coefficient $\alpha$, thickness $d$, refractive indices $n_{\omega}$ and $n_{3 \omega}$, phase mismatch factor $\Delta k$ between the first and third-harmonic wavelength, the laser repetition frequency $f_{\rm rep}$, and the laser pulse duration $\tau_p$. The estimated susceptibility from the above equation is of the order of ($10^{-15}-10^{-19})$ m$^2$/V$^2$ in monolayer graphene~\cite{hendry_PRL2010_coh,kumar_PRB2013_third,Hong2013_thgrphn,Soavi2018_broadbandthg,Jiang2018_gate,Inukai2023_cmplxthg}, ($10^{-17}-10^{-19})$ m$^2$/V$^2$ in black phosphorus~\cite{Youngblood2017_layerthgacs,Zhu2025_bp_mid_infra}, $10^{-19}$ in MoS$_2$~\cite{Woodward2017_microscopy,Saynatjoki2017_thgmos2}. Our estimate of the third-harmonic susceptibility for an $d$-wave altermagnetic phase in RuO$_2$ (see Fig.~\ref{fig4}) is also of the same order of magnitude, and indicates its observability. 

THG has a wide range of applications. Its role in frequency conversion~\cite{Boyd2008_optics_book} often makes it attractive for integrated nonlinear photonics. The strong polarization anisotropy observed in van der Waals layered materials~\cite{Youngblood2017_layerthgacs,Zhu2025_bp_mid_infra} and carbon nanotubes~\cite{Zhu2023_cnt} suggests opportunities for polarization-sensitive devices~\cite{Li2017_nlrev}. In addition, THG microscopy has emerged as a powerful characterization tool: unlike Raman or photoluminescence, it enables rapid, high-contrast imaging of grain boundaries independent of crystal orientation~\cite{Karvonen2017_grain}. These applications highlight the broader relevance and usefulness of the geometric mechanisms for THG identified in our work for fundamental studies as well as for practical applications. 

\section{Conclusion}
\label{disc_sec}

We have developed a comprehensive theory for THG based on the density-matrix formalism, revealing the central role of band geometry in shaping the nonlinear response. By deriving the third-order conductivity tensor, we identified five key band-geometric quantities, quantum metric, Berry curvature, metric and symplectic connections, and second-order connection, that govern intraband and interband processes. Our calculations explicitly incorporate disorder and separate the Fermi sea and Fermi surface contributions. We resolve longstanding discrepancies in graphene’s THG resonances at \(\hbar\omega = 2\mu/3\), \(\mu\), and \(2\mu\), and predict tunable longitudinal and transverse responses in centrosymmetric systems. We explicitly demonstrate this in spin-split altermagnet RuO\(_2\), where the third-order transverse responses reverse under magnetic order. Our symmetry classification across all 122 magnetic point groups further identifies the allowed THG response tensors in each group, to enable targeted material discovery. 

This work lays the foundation for engineering THG in quantum materials, offering tunable responses for THz technologies in telecommunications and ultrafast imaging through doping, gating, or strain. Future theoretical extensions could integrate electron-electron interactions or phonon-assisted processes to refine relaxation dynamics. Additionally, experimental validation via THz spectroscopy probing RuO\(_2\)’s predicted resonances could unlock novel THG devices. 
Our approaches highlight the band geometric contributions to THG responses, motivating further exploration. 

\section{Acknowledgements}
S. S. thanks the MHRD, India, for funding through the Prime Minister’s Research Fellowship (PMRF). D. M thanks IIT Kanpur for funding through the FARE fellowship. A. A. acknowledges funding from the Core Research Grant by Anusandhan National Research Foundation (ANRF, Sanction No. CRG/2023/007003), the Department of Science and Technology of the Government of India, for Project No. DST/NM/TUE/QM-6/2019(G)-IIT Kanpur. Department of Science and Technology, India.  A. A.  acknowledges the high-performance computing facility at IIT Kanpur, including HPC 2013, and Param Sanganak. 
\appendix
\begin{widetext}
\section{Calculation of density matrices }\label{App_A}
In this appendix, we derive the density matrix up to third order in the external electric field by evaluating Eq.~\eqref{dm_nth} for $N=1,2,3.$ 
\subsubsection{First order density matrix}
Setting $N=1$ in Eq.\eqref{dm_nth}, we obtain:
\be
    \rho_{nm}^{(1)}(t)=\frac{i \textit{e}}{\hbar}e^{-i \omega_{nm}t}\int_{-\infty}^{t} dt^\prime e^{i(\omega_{nm}-i/\tau)t^\prime}\bm{E}(t^\prime) \cdot [\bm{R}_{e}^{(0)}+\bm{R}_{i}^{(0)}]~.
\ee
Substituting the equilibrium density matrix $\rho^{(0)}_{nm}$ into this expression, and decomposing the result into intraband and interband parts yields, 
\begin{equation}
    \rho_{nn}^i=-i\frac{\textit{e}}{\hbar \tilde{\omega}}\dfrac{\partial f_n}{\partial k^b}E_b e^{-i\omega t}\label{eq:13}~,
\end{equation}
\be 
 \rho_{nm}^e=\frac{\textit{e}}{\hbar}\frac{\mathcal{R}_{nm}^bf_{mn}}{(\omega_{nm}-\tilde{\omega})}E_b e^{-i\omega t}~,\label{eq:1o}
\ee 
where $\tilde{\omega}=\omega+i/\tau$, and $f_{nm}=f_n-f_m$ is the difference between the occupation of bands $n$ and $m$ in equilibrium. The superscripts $i$ and $e$ indicate contributions arising from $[{\bm r}_i,\rho_0]$ and $[{\bm r}_e,\rho_0]$, respectively. In the above equation, $b$ is summed over. Note that the intraband part of the first-order density matrix is nonzero only in metals with a finite Fermi surface. In contrast, for insulators and semiconductors, only interband contributions are present at first order.
\subsubsection{Second order density matrix}
The second-order density matrix is found by substituting $N=2$
in Eq. \eqref{dm_nth}. It gives
\be
    \rho_{nm}^{(2)}(t)=\frac{i \textit{e}}{\hbar}e^{-i \omega_{nm}t}\int_{-\infty}^{t} dt^\prime e^{i(\omega_{nm}-2i/\tau)t^\prime}\bm{E}(t^\prime)\cdot[\bm{R}_{e}^{(1)}+\bm{R}_{i}^{(1)}]~.\label{eq:r2}
\ee
Here, $\bm{R}_i$ and $\bm{R}_e$ are functions of the first-order density matrix. $\rho^{(2)}$ can be decomposed in four parts, $\rho^{ii}$, $\rho^{ie}$, $\rho^{ei}$, and $\rho^{ee}$, where for example $\rho^{ie}$ results from $[{\bm r}_i,[{\bm r}_e, \rho_0]]$. The four parts of $\rho^{(2)}$ are given by
\be 
\rho^{ii}_{nn}=-\frac{e^2}{2\hbar^2 \tilde{\omega}^2}\frac{\partial^2f_n}{\partial k^c\partial k^b} E_b E_c e^{-i2\omega t}~,
\ee 
\be 
\rho^{ie}_{nm}=i\frac{e^2}{\hbar^2(\omega_{nm}-2\tilde{\omega}) }\left(\frac{\mathcal{R}_{nm}^bf_{mn}}{\omega_{nm}-\tilde{\omega}}\right)_{;k^c}E_b E_c e^{-i2\omega t}~,
\ee 
\be 
\rho^{ei}_{nm}=i\frac{e^2}{\hbar^2 \tilde{\omega}}\frac{\mathcal{R}_{nm}^c}{(\omega_{nm}-2\tilde{\omega})}\frac{\partial f_{nm}}{\partial k^b}E_b E_c e^{-i2\omega t}~,
\ee 
and 
\be
\rho^{ee}_{nm}=\frac{e^2}{\hbar^2(\omega_{nm}-2\tilde{\omega})}\sum_l\left(\frac{\mathcal{R}_{nl}^c \mathcal{R}_{lm}^bf_{ml}}{\omega_{lm}-\tilde{\omega}}- \frac{\mathcal{R}_{nl}^b \mathcal{R}_{lm}^cf_{ln}}{\omega_{nl}-\tilde{\omega}}\right)E_b E_c e^{-i2\omega t}~.
\ee 
Note that $\rho^{ii}$ and $\rho^{ei}$ are finite only for metals having a finite Fermi surface. On the other hand, $\rho^{ie}$ and $\rho^{ee}$ contribute to the nonlinear optical response even in insulating systems at $T = 0$. 
\subsubsection{Third-order density matrix}
To calculate the third-order density matrix we need to use $N=3$ in Eq. \eqref{dm_nth}, to obtain 
\be
    \rho_{nm}^{(3)}(t)=\frac{i \textit{e}}{\hbar}e^{-i \omega_{nm}t}\int_{-\infty}^{t} dt^\prime e^{i(\omega_{nm}-3i/\tau)t^\prime}\bm{E}(t^\prime)\cdot[\bm{R}_{e}^{(2)}+\bm{R}_{i}^{(2)}]~.\label{eq:r2}
\ee
Here, $\bm{R}_i^{(2)}$ and $\bm{R}_e^{(2)}$ are functions of the second-order density matrix. $\rho^{(3)}$ can be expressed as a sum of eight parts, $\rho^{iii}$, $\rho^{eii}$, $\rho^{iei}$, $\rho^{eei}$, $\rho^{iie}$, $\rho^{eie}$, $\rho^{iee}$, and $\rho^{eee}$. The origin of each term having a given superscript can be traced to the corresponding commutator. For example $\rho^{eie}$ originates from $[{\bm r}_e,[{\bm r}_i,[{\bm r}_e, \rho_0]]]$. The eight parts of $\rho^{(3)}$ are given by
\be
\rho^{iii}_{nn}=i\frac{e^3}{6\hbar^3 \tilde{\omega}^3}\frac{\partial^3f_n}{\partial k^d\partial k^c\partial k^b}E_b E_c E_d e^{-i3\omega t}~,
\ee
\be 
\rho^{eii}_{nm}=-\frac{e^3}{2\hbar^3 \tilde{\omega}^2}\frac{\mathcal{R}_{nm}^d}{(\omega_{nm} - 3\tilde{\omega})}\frac{\partial^2f_{nm}}{\partial k^c\partial k^b} E_b E_c E_d e^{-i3\omega t}~,
\ee
\be
\rho^{iei}_{nm}=-\frac{e^3}{\hbar^3 \tilde{\omega}(\omega_{nm} - 3\tilde{\omega})} \left[\frac{\mathcal{R}_{nm}^c}{(\omega_{nm} - 2\tilde{\omega})}\frac{\partial f_{nm}}{\partial k^b}\right]_{;k^d}E_b E_c E_d e^{-i3\omega t}~,
\ee
\be
\rho^{eei}_{nm}=i\frac{e^3}{\hbar^3 \tilde{\omega}(\omega_{nm} - 3\tilde{\omega})}\sum_p\left[\frac{\mathcal{R}_{np}^d \mathcal{R}_{pm}^c}{(\omega_{pm}-2\tilde{\omega})}\frac{\partial f_{pm}}{\partial k^b} - \frac{ \mathcal{R}_{np}^c \mathcal{R}_{pm}^d }{(\omega_{np}-2\tilde{\omega})} \frac{\partial f_{np}}{\partial k^b}\right]E_b E_c E_d e^{-i3\omega t}~,
\ee
\be
\rho^{iie}_{nm}=-\frac{e^3}{\hbar^3(\omega_{nm} - 3\tilde{\omega})}\left[\frac{1}{(\omega_{nm} - 2\tilde{\omega})} \left(\frac{\mathcal{R}_{nm}^bf_{mn}}{\omega_{nm}-\tilde{\omega}}\right)_{;k^c}\right]_{;k^d}E_b E_c E_d e^{-i3\omega t}~,
\ee 
\be
\rho^{eie}_{nm}=-i\frac{e^3}{\hbar^3(\omega_{nm} - 3\tilde{\omega})}\sum_p\left[\left(\frac{\mathcal{R}_{np}^bf_{pn}}{\omega_{np}-\tilde{\omega}}\right)_{;k^c}\frac{\mathcal{R}_{pm}^d}{(\omega_{np}-2\tilde{\omega})} - \frac{\mathcal{R}_{np}^d}{(\omega_{pm}-2\tilde{\omega})}\left(\frac{\mathcal{R}_{pm}^bf_{mp}}{\omega_{pm}-\tilde{\omega}}\right)_{;k^c} \right]E_b E_c E_d e^{-i3\omega t}~,
\ee
\be
\rho^{iee}_{nm}=i\frac{e^3}{\hbar^3(\omega_{nm} - 3\tilde{\omega})}\left[\sum_p \frac{1}{(\omega_{nm}-2\tilde{\omega})}\left(\frac{\mathcal{R}_{np}^c \mathcal{R}_{pm}^bf_{mp}}{\omega_{pm}-\tilde{\omega}} - \frac{\mathcal{R}_{np}^b \mathcal{R}_{pm}^cf_{pn}}{\omega_{np}-\tilde{\omega}}\right)\right]_{;k^d} E_b E_c E_d e^{-i3\omega t}~,
\ee
and 
\bea
\rho^{eee}_{nm}&=&\frac{e^3}{\hbar^3(\omega_{nm} - 3\tilde{\omega})}\sum_{p,l}\left[\frac{\mathcal{R}_{np}^d}{(\omega_{pm}-2\tilde{\omega})}\left(\frac{\mathcal{R}_{pl}^c \mathcal{R}_{lm}^bf_{ml}}{\omega_{lm}-\tilde{\omega}} - \frac{\mathcal{R}_{pl}^b \mathcal{R}_{lm}^cf_{lp}}{\omega_{pl}-\tilde{\omega}}\right) - \left(\frac{\mathcal{R}_{nl}^c \mathcal{R}_{lp}^bf_{pl}}{\omega_{lp}-\tilde{\omega}}- \frac{\mathcal{R}_{nl}^b \mathcal{R}_{lp}^cf_{ln}}{\omega_{nl}-\tilde{\omega}}\right)\frac{\mathcal{R}_{pm}^d}{(\omega_{np}-2\tilde{\omega})}\right] \nn\\&\times&E_b E_c E_d e^{-i3\omega t}~.
\eea 
We note that the first three components are finite for metals with a definite Fermi surface. On the other hand, the remaining terms can be finite even in insulators and semiconductors.  This completes our calculation of density matrices up to the third-order in the electric field. 

\end{widetext}
\bibliography{THG.bib}
\end{document}